# Anomalous two peak structure in the Angle-resolved Photoemission Spectra of $Ba_{1-x}K_xFe_2As_2$

T. Shimojima[1,2], W. Malaeb[3,4], K. Ohgushi[3,4], A. Chainani[5], S. Shin[2,3,4,5],
S. Ishida[6,7], M. Nakajima[6,7], S. Uchida[4,7], T. Saito[8], H. Fukazawa[4,8], Y. Kohori[4,8], K. Kihou[4,6],
C.H. Lee[4,6], A. Iyo[4,6], H. Eisaki[4,6], K. Ishizaka[1,2]

[1] *Department of Applied Physics, University of Tokyo, Tokyo 113-8656, Japan*

[2] *CREST, JST, Chiyoda-ku, Tokyo 102-0075, Japan*

[3] *Institute for Solid State Physics (ISSP), University of Tokyo, Kashiwa, Chiba 277-8581, Japan*

[4] *TRIP, JST, Chiyoda-ku, Tokyo 102-0075, Japan*

[5] *The Institute of Physical and Chemical Research (RIKEN), Sayo-gun, Hyogo 679-5148, Japan*

[6] *National Institute of Advanced Industrial Science and Technology (AIST), Tsukuba, Ibaraki 305-8568, Japan*

[7] *Department of Physics, University of Tokyo, Tokyo 113-8656, Japan*

[8] *Department of Physics, Chiba University, Chiba 263-8522, Japan*

The electronic structure near the Fermi level ($E_F$) of $Ba_{1-x}K_xFe_2As_2$ (BaK122 ; $x = 0.2$ - 0.7) is studied using laser ultrahigh-resolution angle-resolved photoemission spectroscopy(ARPES). For the optimally doped case of $x = 0.4$, we clearly observe two peaks below $T_c$ in the ARPES spectra at a binding energies (BE) of 5 meV and 13meV. The former is assigned to a superconducting (SC) coherence peak since it appears and evolves below the bulk SC transition at $T_c$ (= 36 K), accompanying a gap opening centered at $E_F$. In contrast, the latter peak, which appears below ~ 90 K without any gap formation, is interpreted to be not directly related to a SC coherence peak. This high-BE peak is observed from $x = 0.2$ to 0.6, reduces in energy with overdoping ($x > 0.4$) and is absent for $x = 0.7$. The temperature($T$)- and doping-dependent ARPES results suggest that the high-BE peak originates from coupling to a bosonic mode of energy $\Omega$ ~ 8 meV.

## 1. Introduction

The parent compounds of Iron-pnictide superconductors[1] typically show antiferromagnetic (AF) metallic ground states. Introducing carriers, isovalent-ions or applying pressure causes superconductivity in the vicinity of the magnetic and structural phase boundaries. In the SC state, electrons form the Cooper pairs across and/or within the disconnected quasi-two-dimensional Fermi surfaces (FSs) composed of the Fe $3d$ multi-orbital band structure [2], resulting in the SC gap formation at the $E_F$. Nature of the pairing force can be deduced from the symmetry of the SC gaps. ARPES has been an important probe to directly elucidate the SC gap variations in the momentum space. Despite the intensive research by ARPES to date, nevertheless, the electron pairing mechanism in the iron-based superconductors remains to be clarified.

One of the iron-pnictides most investigated by ARPES is the hole-doped BaK122 system[3]. Optimal doped BaK122 shows three hole-FSs (inner, middle and outer-FSs) located around the Brillouin zone center and one electron FS at the zone corner[4]. In the early stage, SC gaps without nodes in the entire FSs had been reported for optimal doped BaK122[5-8]. The inner hole- and the electron-FS sheets showed a larger SC gaps of ~ 12 meV, while the outer hole FS exhibited a smaller SC gap of ~ 4 meV. This result is what expected by a simple spin fluctuation mechanism based on the FS nesting[9,10]. In this picture, the smaller SC gap in the outer hole FS is associated with the insufficient nesting condition which weakens the superconductivity.

On the other hand, a different result is reported by a recent laser-ARPES study[11]; three hole-FSs around the zone center showing comparable SC gap sizes (~ 3 meV) near $k_z = \pi$. This observation was interpreted as an indication of the inter-orbital pair scattering via orbital fluctuations[12,13]. This experimental discrepancy in the FS sheet dependence of the SC gaps is mainly caused by the complex spectral feature in the energy distribution curves (EDCs) in BaK122 system. Several groups observed a two-peak or peak-with-shoulder structure

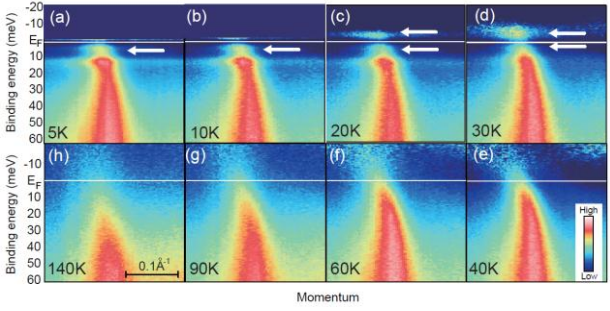

**Fig.1** (a-d) ARPES images divided by FD function measured by p-polarization in the SC state. White arrow in (a) - (d) indicates appearance of the SC coherence peak below and above $E_F$. (e-h) ARPES images divided by FD function measured by p-polarization in the normal state.

by ARPES using different photon energies[5,6,11,14,15]. A careful separation and assignment of the origin of these two peaks are thus required for investigating the SC gap symmetry in BaK122 system.

In this paper, we report the detailed $T$-dependent and $x$-dependent laser-ARPES on $Ba_{1-x}K_xFe_2As_2$ system (BaK122 ; $x$ = 0.2 - 0.7). Recent development of the laser-ARPES technique, especially in the energy resolution and bulk sensitivity, enables us to investigate the fine structure of the EDCs. The laser ARPES probes the in-plane electronic structure near $k_z = \pi$, and we could clearly separate out the two peaks located in the EDCs at the BE of 5 meV and 13 meV, respectively. The former is assigned to a SC coherence peak since it appears and evolves below bulk $T_c$ = 36 K accompanying a gap opening centered at $E_F$. In contrast, the latter peak is interpreted to be not directly related to SC coherence because of its existence up to ~ 90 K, i.e. far above bulk $T_c$ and without a gap formation. The high-BE peak is observed from $x$ = 0.2 to 0.6 with smooth variation in its energy, but is not observed for $x$ = 0.7. Evolution of the high-BE peak associated with the formation of the kink structure in the band dispersion. $T$- and $x$-dependent ARPES imply that the appearance of the high-BE peak is linked to the coupling with a bosonic mode of energy $\Omega$ ~ 8 meV.

## 2. Experimental

High-quality single crystals of under-doped (UD) ($x$ =0.2, 0.3, 0.4) and over-doped (OD) ($x$ = 0.5, 0.6, 0.7) BaK122 were grown by using the FeAs[17] and KAs [18] flux, respectively.

Laser-ARPES measurements were performed on a spectrometer built using a GAMMADATA-SCIENTA R4000WAL electron analyzer and an ultraviolet laser (hν = 6.994 eV) as a photon source[16]. The energy resolution was set to 3 meV in order to get enough photoelectron count rate. Single crystals of BaK122 are mounted on a copper plate using a silver paste. $E_F$ is determined by measuring the gold film evaporated near the sample with an accuracy of better than ± 0.1 meV.

## 3. Two-peak structure at low temperature

Optimally doped BaK122 with $x$ = 0.4 shows three hole FSs around the zone center at $k_z = \pi$. The inner and middle FSs are almost degenerate near $k_z = 0$[4]. In order to separately observe the inner-hole band, it was necessary to choose $p$-polarization of the incident photons. Fig.1 (a-h) show $E$-$k$ images measured along $(0,0) - (\pi,\pi)$ line. All $E$-$k$ images are divided by the Fermi Dirac (FD) function in order to emphasize the unoccupied state within ~ $5k_BT$ of $E_F$. Below bulk $T_c$ = 36 K [Fig.1(a-d)], intensity near $E_F$ rapidly decreases within an energy window of +/-5meV and peaks appear around ~ 5 meV below and above $E_F$ [white arrows in Fig.1 (a-d)]. At low-$T$, we also find another structure around 13 meV, which has non-dispersive tails in a wide momentum region (Fig1.(a-d)).

In order to closely investigate the $T$-dependence, we plot the EDCs at $k_F$ of the inner hole band as a function of T from 10 to 170 K(Fig.2 (a-c)) and those divided by FD function (Fig.2 (d-f)). As seen in Fig. 2 (d) and (e), a gap opening below bulk $T_c$ is observed in the energy window of ±5 meV centered at $E_F$. This observation indicates that a pair of peaks at BE of +5 meV and -5 meV corresponds to the quasiparticle peaks directly related to the superconducting coherence peak.

Here we quantitatively estimate the SC gap magnitude. Referring to the SC peak position, we can estimate the SC gap of ~ 5 meV. In this case, $2\Delta/k_BT_c$ becomes ~ 2.8. However, we note that the SC gap size estimated in this way tends to be larger than values obtained by fitting the spectra, particularly for the case of rather broad SC peaks as observed for the iron-pnictides. Instead, by fitting the EDC with BCS spectral functions in a similar manner as was used in reference 11, we obtained the SC gap magnitude of ~ 3 meV corresponding to a $2\Delta/k_BT_c$ ~ 1.7. This is significantly smaller than the BCS weak coupling limit of $2\Delta/k_BT_c$ = 3.52.

Such a small gap value has been theoretically shown for the case of multi-gap superconductivity with the electron pairing between disconnected FSs[19]. In this picture, the small density of states (DOS) at one FS reflects itself in the small SC gap magnitude at the other FS, regardless of the nature of the pairing glue. According to the band calculations, DOS of each of the hole-FSs is larger than that of each electron-FSs in BaK122[20]. Small SC gap in the hole-FSs observed by laser-ARPES thus suggests the electron pair scattering between the disconnected hole- and electron-FSs.

This also implies that the electron-bands have a larger $2\Delta/k_BT_c$ ratio than hole-bands, which can account for the point contact Andreev reflection [21] and the lower critical field[22] measurements reporting two-gap superconductivity in BaK122, where the smaller gap size actually corresponds well with that obtained by laser-ARPES. We actually observed a larger SC gap of ~ 10 meV, without a two-peak structure, in the electron FS of the same single crystal of optimal doping using a synchrotron photon source [23]. This result also supports our interpretation of the two-gap picture in BaK122.

Next we move on to the interpretation of the origin of the high-BE peak in the hole FSs. In the normal state, EDCs show asymmetric spectral weight across $E_F$ [Figs.2 (c,f)]. Furthermore, intensity at $E_F$ significantly increases on lowering $T$. The hump structure starts to evolve around 90 K only in the occupied state at ~ 8 meV. Based on the continuous increase of the intensity at $E_F$, we conclude that the hump structure does not accompany any gap structure formation. We can thus mention that this $T$-dependent hump structure, i.e. *pseudo-peak*, should be distinguished from the *pseudo-gap* phenomena observed in cuprates[24,25].

Similar $T$-dependent evolution of the peak has been observed in two-dimensional systems such as $Na_xCoO_2$, across a characteristic $T$ where dimensionality (D) in the transport properties change from 2-D to 3-D[26]. Another case is $LiV_2O_4$ showing a Kondo-resonance peak just above $E_F$ across the crossover into the heavy-fermion-like state [27]. In the section 5, we will discuss the relevance between transport properties and the pseudopeak evolution in BaK122.

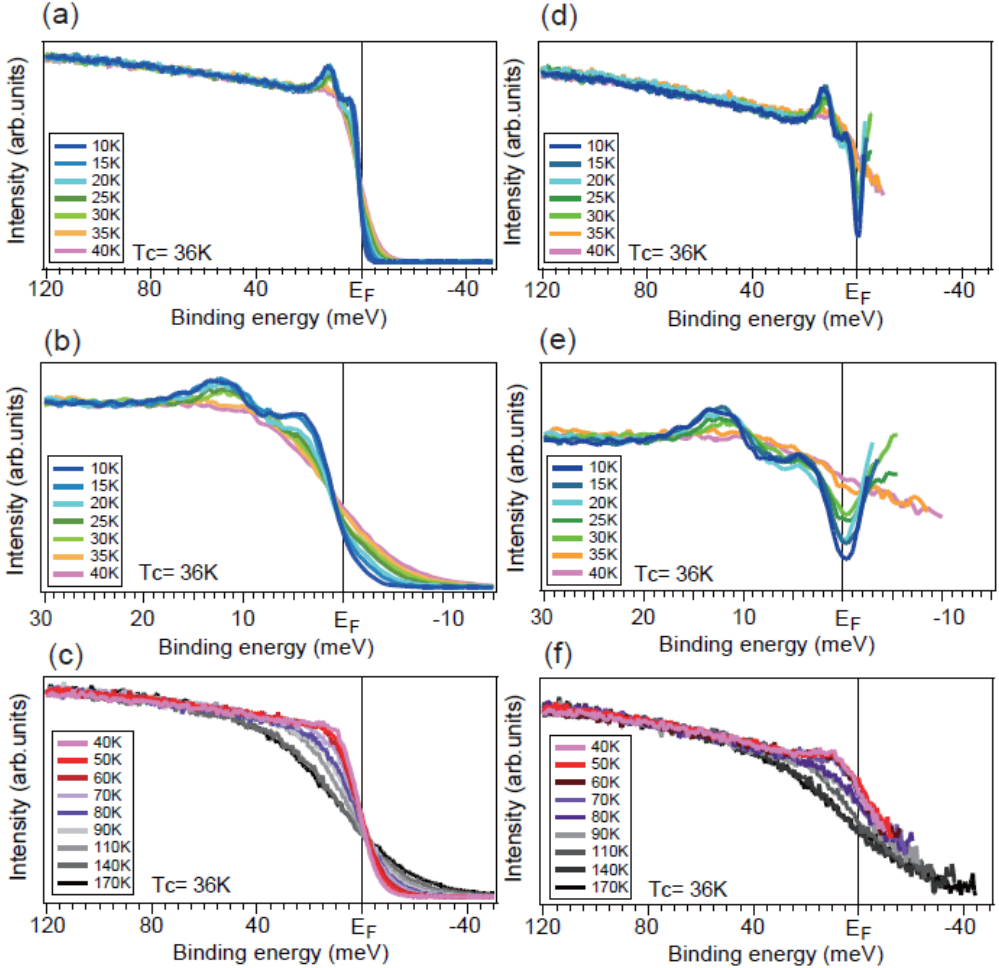

**Fig.2** (a) Energy distribution curves of inner-hole band at 10 – 40 K for BaK122 at $x$ = 0.4. (b) Energy distribution curves at 10 – 40 K. in an enlarged energy scale. (c) Energy distribution curves of inner-hole band at 40 – 170 K. (d-f) Energy distribution curves in (a-c) divided by Fermi Dirac function

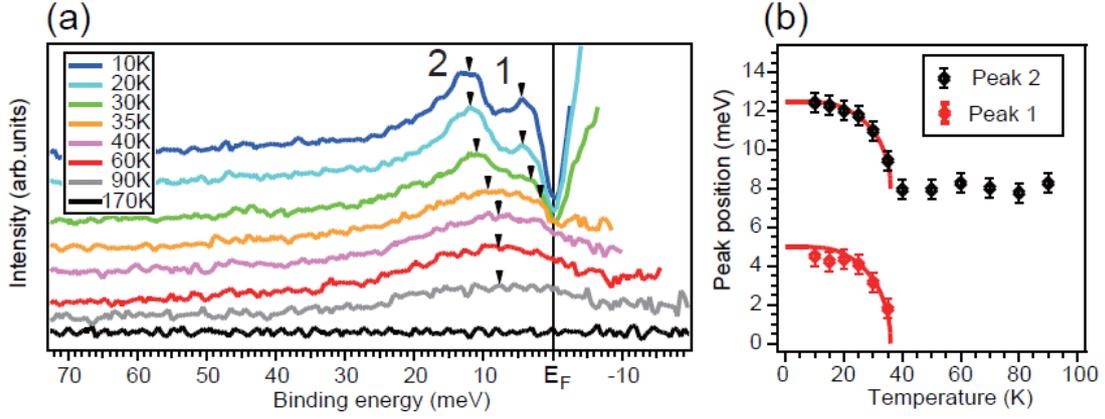

**Fig.3** (a) Energy distribution curves of inner hole band at 10 – 170 K divided by Fermi Dirac function and further normalized by that of 170 K .for BaK122 at $x$ = 0.4. (b) $T$-dependence of the energy position of peak 2 (pseudopeak) and peak 1 (superconducting peak).

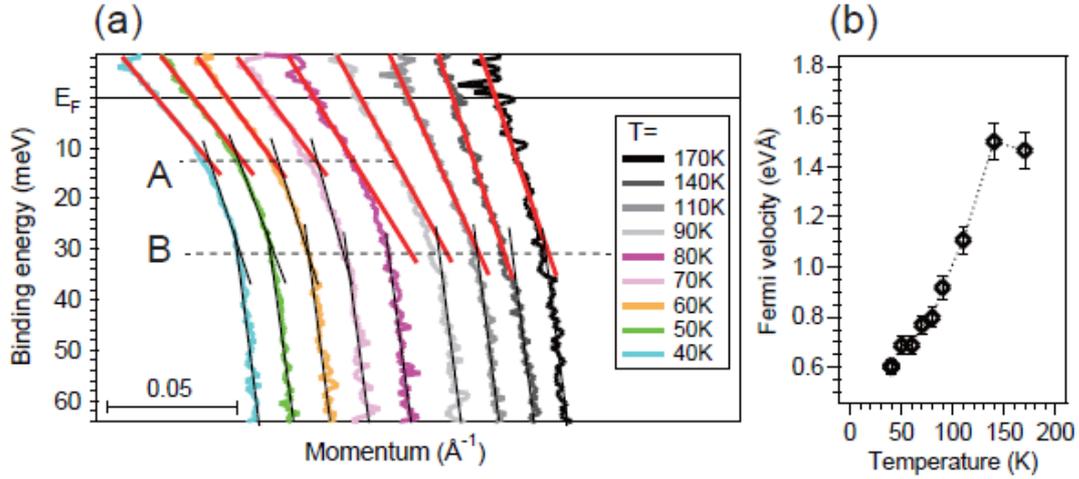

**Fig.4** (a) Band dispersion of inner hole band measured at 40 K– 170 K for BaK122 at $x$ = 0.4. Temperature-dependent part of the band dispersions are guides for eyes (colored in red). A and B represents the two energy scales of the kink structures determined from the crossing point of guides for eyes. (b) Temperature-dependence of the Fermi velocity.

Here we comment on the consistency of our data in comparison with the previous ARPES studies using different photon sources. In the previous reports[5,14,15], similar two-peak or peak-with-shoulder structure had been observed in EDCs of BaK122. Despite the different photon energies[5,11,15], i.e. different $k_z$ values, the energy positions of the two peaks are basically identical, thereby indicating that the present data itself is consistent with those of the previous reports.

On the other hand, what is different from the previous ARPES reports is the interpretation regarding the origin of the two peaks. References [5,6,12] concluded that the high-BE peak in the inner-hole band was a SC peak resulting in the SC gap $\Delta_{inner}$ of ~ 12 meV estimated from the peak position. In addition, low-BE peak was discussed as an extrinsic spectral feature due to the non-SC signal from the aged surface[14].

However, a recent bulk sensitive and

high energy resolution laser-ARPES study reveals that the low-BE peak rather reflects the bulk superconductivity whereas the high-BE peak shows the $T$-dependent evolution starting from much above $T_c$. We believe that both the peaks originate in the bulk physical properties of the BaK122 system, as will be discussed in detail in the following sections 4 and 5.

We next move onto the detailed $T$-dependence across $T_c$ in order to get a deeper insight of the origin of the pseudopeak. Figure 3 (a) shows the $T$-dependence across $T_c$ of the EDCs near $E_F$ for inner hole-band. All spectra are divided by the FD function and further normalized by the spectrum at 170 K. Black bars indicate the peak position of each peak. While the SC peak (peak 1) appears below $T_c$ and follows the BCS-like $T$-evolution, the pseudo-peak (peak 2) slightly shifts in energy toward higher BE, as shown in Figs. 3(a) and (b). Interestingly, the energy shift of the pseudo-peak corresponds to that of the SC peak. This observation is understood by considering the coupling to a bosonic mode of energy $\Omega \sim 8$ meV. According to the model used in ref. [28], a coupling with an Einstein mode of energy $\Omega$ in the SC state will be observed at a B.E. of $\Omega + \Delta(T)$, where $\Delta(T)$ is the SC gap magnitude. This explains the energy shift of $\Delta(T)$ in the pseudopeak position across $T_c$ as shown in Fig.3(a,b).

Such a mode coupling causing the peaudopeak might be reflected in the renormalization in the quasiparticle properties. In Fig. 4(a), we show the $T$-dependence of the band dispersions obtained from the peak position of the momentum distribution curves (MDC) in the normal state. One can see the kink structure around 30 meV, being consistent with the observation in ref [29]. In the $T$ region ranging from 40 K to 170 K, the kink structure around 30 meV seems almost $T$-independent, implying the contribution of the phonons. Actually, LaFeAsO$_{0.9}$F$_{0.1}$ shows a large weight of phonon density of states around 30 meV[30]. In contrast, a strongly $T$-dependent kink structure is observed at around 10 meV. It appears below ~ 100 K, accompanying the continuous decrease of the Fermi velocity $v_{F\_h}$ as shown in Fig.4.(b). Taking into account of its $T$ range and energy scale, this kink structure should correspond to the pseudopeak formation. Observation of the kink structure at around ~ 10 meV supports the existence of the collective mode energy $\Omega \sim 8$ meV.

**4. Composition dependence of pseudopeak**

We show the $x$-dependence of the two peak structure in the EDCs in Fig. 5 (a). Pseudopeak appears in a wide range of K-doping level from $x = 0.2$ to 0.6, showing a smooth variation with $x$ and a maximum in energy around $x = 0.4$. Such a systematic $x$-dependent variation of the pseudopeak suggests that the pseudopeak is not a surface-derived extrinsic feature. We note that the diminishing nature of the pseudopeak around $x > 0.6$ corresponds to the disappearance of the electron-FS at zone corners, induced by hole doping [31]. This suggests that the origin of the pseudopeak is relevant to the electron scattering between the hole- and electron-FSs. Theoretically, both spin fluctuation and orbital fluctuation are the candidates that enable the electron scattering with a $(\pi,\pi)$ wave vector[12].

Recent neutron scattering measurements reported that a crossover from commensurate to incommensurate electron scattering occurs across $x \sim 0.5$[32]. This suggests that since the electron-FS at X point disappears by hole doping, the flower-like hole-FS near X point starts to play a more important role for the electron scattering in the over-doped region. This is consistent with the disappearance of the electron-FS across $x \sim 0.55$ observed by ARPES[31]. Furthermore, the magnetic excitation energy shows a $x$-dependence similar to that of the pseudopeak energy[32]. These results imply that spin fluctuations is a probable candidate for a bosonic mode of $\Omega \sim 8$ meV.

Figure 5(c) shows the FS sheet dependence of the EDCs measured by $p$-polarization (inner-FS) and $s$-polarizations (middle- and outer-FSs), respectively. It is remarkable that the positions of the two peaks (marked as 1 and 2) are FS sheet-independent. This indicates that the bosonic mode with 8 meV couples with electrons irrespective of orbital character. This observation, however, seems to go against the spin fluctuation scenario since spin fluctuations are enhanced only among the FS sheets showing good nesting condition. In order

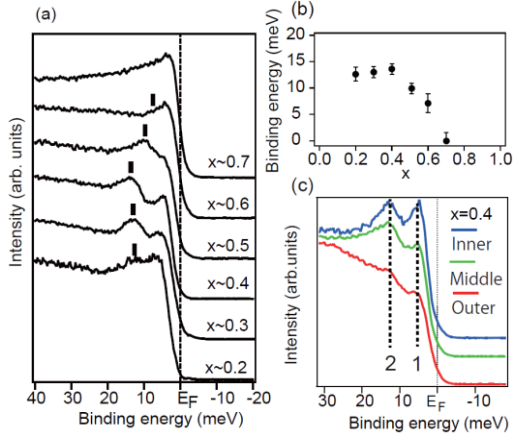

**Fig.5** (a) Energy distribution curves of inner-hole band at 10 K for $x = 0.2 – 0.7$. (b) Binding energy of pseudopeak for various compositions determined by the black bars in (a). (c) Energy distribution curves of inner- (blue), middle- (green) and outer- (red) hole bands at $x = 0.4$.

to conclusively identify the origin of pseudopeak, further experiments on its $T$-dependence in a wide range of $x$ will be required.

**5. Comparison with transport measurements**

Here we compare the ARPES results with transport measurements. Magnetoresistance is known to show a non-monotonic $x$-dependent anomalous feature at 80 – 100 K, from under doped- to optimal-doped region[33]. On the other hand, the $T$-dependence of the Hall coefficient $|R_H|$ has been reported[17] to show a reduction below ~ 100 K in $0.26 < x < 0.55$. One can infer that the anomaly and its doping dependence in both Hall coefficient and magnetoresistance are possibly related to the appearance of the pseudopeak. Adopting the Boltzmann equation for a two-band system, the decrease in $|R_H|$ can be connected to the increase of $m_h/m_e$[17]. Increase in $m_h$ (or the decrease in $v_{F\_h}$) below ~ 100 K accompanied by the pseudopeak formation observed by ARPES for the optimally doped sample (Fig.4 a, b), may support this interpretation. The fact that the peaudopeak has been observed in the $x$- and $T$- region where the abovementioned transport properties show the anomalies strongly suggests that the pseudopeak originates in the bulk physical property. In order to obtain a conclusive answer for the origin of pseudopeak, further investigations of precise measurements of transport, thermal and magnetic properties in comparison with ARPES are highly required.

**6. Conclusion**

In this work, we investigated the two-peak structure in the EDC of BaK122 using the laser-ARPES. Owing to the high-resolution and bulk-sensitivity, two peaks were clearly observed separately below $T_c$ at the BE of 5 meV and 13 meV. According to the detailed $T$-dependent ARPES, the former is assigned to a SC coherence peak since it appears and evolves below bulk $T_c = 36$ K, accompanying a gap opening centered at $E_F$. In contrast, the latter cannot be directly interpreted as a SC coherence peak because of its existence up to ~ 90 K without any (pseudo) gap structure. The high-BE peak was observed from $x = 0.2$ to 0.6 with a systematic variation in its energy. Evolution of the high-BE peak is associated with the formation of the kink structure in the band dispersion. $T$-dependent and $x$-dependent ARPES implies that the pseudopeak originates in the coupling to a bosonic mode of ~ 8 meV.

**Acknowledgement:** We acknowledge R. Arita, H. Ikeda, K. Kuroki and H. Kontani for valuable discussions. This research is supported by the Japan Society for the Promotion of Science (JSPS) through its FIRST Program.